\documentclass[a4paper]{llncs}
\providecommand{\url}[1]{#1}
\usepackage{amssymb,amsmath,enumerate,graphicx}

\usepackage{color}

\usepackage{tikz}
\usetikzlibrary{snakes}
\tikzstyle{hyb}=[rectangle,fill=green!50,draw,minimum size=3.5mm]
\tikzstyle{tre}=[circle,fill=green!50,draw,minimum size=3.5mm]
\newcommand{\etq}[1]{%
\draw (#1) node {\footnotesize $#1$};
}

\newcommand{\contained}{\preccurlyeq}

\newcommand{\pathgr}{\!\rightsquigarrow\!{}}

\renewcommand{\leq}{\leqslant}
\renewcommand{\geq}{\geqslant}

\newcommand{\NN}{\mathbb{N}}
\newcommand{\RR}{\mathbb{R}}

\begin{document}

\title{On Nakhleh's latest metric for phylogenetic networks}

\author{Gabriel Cardona\inst{1} \and Merc\`e Llabr\'es\inst{1} \and Francesc Rossell\'o\inst{1} \and
Gabriel Valiente\inst{2,3} } \authorrunning{G. Cardona et al.}
\institute{Department of Mathematics and Computer Science, University
of the Balearic Islands, E-07122 Palma de Mallorca,
\{\texttt{gabriel.cardona,merce.llabres,cesc.rossello}\}\texttt{@uib.es}
 \and
Algorithms, Bioinformatics, Complexity and Formal Methods Research
Group, Technical University of Catalonia, E-08034 Barcelona,
\texttt{valiente@lsi.upc.edu} }

\maketitle

\begin{abstract}
We prove that Nakhleh's latest   `metric' for phylogenetic networks  is a metric on the classes of tree-child phylogenetic networks, of semi-binary time consistent tree-sibling phylogenetic networks, and of multi-labeled phylogenetic trees.We also prove that it  separates distinguishable phylogenetic networks. In this way, it becomes the strongest dissimilarity measure for phylogenetic networks available so far. \end{abstract}
%

\section{Introduction}
\label{sec:intro}

Evolutionary networks are explicit models of evolutionary histories that include reticulate evolutionary events like genetic recombinations, lateral gene transfers or hybridizations. There are currently many algorithms and software tools that make it possible to reconstruct evolutionary networks. As in the classical phylogenetic tree reconstruction setting, the assessment of evolutionary network reconstruction methods requires the ability to compare phylogenetic networks; for instance, to compare inferred networks with either simulated networks or true phylogenies, and to evaluate the robustness of phylogenetic network reconstruction algorithms when adding new species~\cite{moret.ea:2004,woolley.ea:2008}. This has led to an increasing interest in the defintion of dissimilarity measures for the comparison of evolutionary networks, and their implementation in software packages.

These  dissimilarity measures include the \emph{bipartitions}, or \emph{Robinson-Foulds}, metric \cite{baroni.ea:ac04}, which satisfies the axioms of metrics \cite{deza2006dd}  on the classes of {regular networks}  \cite{baroni.ea:ac04} and of tree-child time consistent phylogenetic networks \cite{cardona.ea:07a}; the \emph{tripartitions} metric \cite{moret.ea:2004}, which satisfies the axioms of metrics on the class of
tree-child time consistent phylogenetic networks \cite{cardona.ea:07a}; the \emph{$\mu$-distance} \cite{cardona.ea:07b}, which is a metric on the classes of all tree-child phylogenetic networks \cite{cardona.ea:07b} and of semi-binary tree-sibling time consistent phylogenetic networks \cite{cardona.ea:sbTSTC:2008}; the \emph{triplets metric}, which is a metric on the class of 
tree-child time consistent phylogenetic networks \cite{comparison2}; and a \emph{nodal metric} that is again a metric on the class of 
tree-child time consistent phylogenetic networks \cite{comparison2,cardona.ea:08nn}.

L. Nakhleh has recently proposed a dissimilarity measure for the comparison of
phylogenetic networks \cite{nakhleh:2007} and he has proved that
it satisfies the separation axiom of metrics (zero distance means
isomorphism) on the class of all \emph{reduced} phylogenetic networks
in the sense of \cite{moret.ea:2004}, and hence that it is a metric on this class of networks.  
In this note (which should be seen as a sequel of our previous technical report \cite{cardona.ea:08-2metrics}) we complement and generalize Nakhleh's work in two directions.  On the
one hand, we prove a stronger result: namely, that  for this dissimilarity measure, zero distance implies
indistinguishability up to reduction in the sense of
\cite{moret.ea:2004}, a goal that had already been unsuccessfully pursued by
Moret-Nakhleh-Warnow et al in \textsl{loc. cit.}.  In this way, and to the best of our
knowledge,  Nakhleh's dissimilarity measure turns out to be the first one  that separates distinguishable networks. On the other hand, we show that this dissimilarity measure is a metric on several classes of phylogenetic networks and related (multi-)labelled DAGs: namely, 
on the classes of tree-child phylogenetic networks, of semi-binary time consistent tree-sibling phylogenetic networks, and of multi-labelled phylogenetic trees.
Adding this to the aforementioned fact, previously proved by Nakhleh, that it is a metric on the class of all reduced networks, it turns out that his latest dissimilarity measure for phylogenetic networks has the strongest separation power among all metrics defined so far.

\section{Preliminaries}

\subsection{Notations}
\label{sec:prel}

Let $N=(V,E)$ be a DAG (a finite directed acyclic graph).  We say that a node
$v\in V$ is a \emph{child} of $u\in V$ if $(u,v)\in E$; we also say
then that $u$ is a \emph{parent} of $v$. Two children of a same parent are said to be \emph{sibling} of each other. A \emph{leaf} is a node without children.  A node that is not a leaf is called
\emph{internal}. 
 We say that a node is a
\emph{tree node} when it has at most one parent, and that it is a \emph{hybrid node} when it
has more than one parent.  A DAG  is  \emph{quasi-binary} when all its hybrid nodes have exactly two parents and one child, without any restriction on the number of children of its tree nodes.  A DAG is \emph{rooted} when it has only one
\emph{root}: a node without parents. 

A \emph{path} in $N=(V,E)$ is a sequence of nodes $(v_0,v_1,\dots,v_k)$ such
that $(v_{i-1},v_{i})\in E$ for all $i=1,\dots,k$.  We call $v_{0}$
the \emph{origin} of the path, $v_1,\ldots,v_{k-1}$ its \emph{intermediate nodes}, $v_{k}$ its \emph{end}, and $k$ its
\emph{length}; a path is \emph{non-trivial} when its length is larger than 0.  We denote by $u\pathgr v$ any path with origin $u$ and
end $v$ and, whenever there exists a path $u\pathgr v$, we say that
$v$ is a \emph{descendant} of $u$ and that $u$ is an \emph{ancestor} of $v$: if  the path $u\pathgr v$ is non-trivial, we say that $v$ is a  \emph{proper} descendant of $u$ and
that $u$ is an \emph{proper}  {ancestor} of $v$.

The \emph{height} $h(v)$ of a node $v$ in a DAG $N$ is the largest
length of a path from $v$ to a leaf.  The absence of cycles implies
that the nodes of a DAG can be stratified by means of their heights:
the leaves are the nodes of height 0 and, for every $m\geq 1$, the nodes of height $m$ are those internal nodes   with all their children of height smaller
than $m$ and at least one child of height exactly $m-1$.

Let $S$ be a non-empty finite set, whose elements are called \emph{taxa} or other Operational Taxonomic Units;
unless otherwise stated,
for simplicity we shall always take as  $S$  the set
of positive integers $\{1,\ldots,n\}$, with $n=|S|$.
 A \emph{phylogenetic network} on a set $S$ of taxa is a rooted DAG. whose leaves are bijectively labeled by elements of
$S$.  A \emph{phylogenetic tree} is a phylogenetic network without hybrid nodes. We shall always identify, usually without any further notice,
each leaf of a phylogenetic network with its label.  Two phylogenetic networks $N,N'$   are \emph{isomorphic}, in
symbols $N\cong N'$, when they are isomorphic as directed graphs and
the isomorphism sends each leaf of $N$ to the leaf with the same label in $N'$.

A phylogenetic network $N=(V,E)$ is said to be \emph{tree-child} when every internal node has some child that is a tree node,  \emph{tree-sibling} when every hybrid node has some sibling that is a tree node, and  \emph{time consistent} when 
 it allows a mapping
$$
\tau:V\to \NN
$$
such that, for every arc $(u,v)\in E$, $\tau(u)<\tau(v)$ if $v$ is a tree node and
$\tau(u)=\tau(v)$ if $v$ is a hybrid node.
The biological meaning of these conditions has been discussed in 
\cite{baroni.ea:sb06,cardona.ea:sbTSTC:2008,cardona.ea:07b,moret.ea:2004}.

For every node $u$ of a phylogenetic network $N=(V,E)$, let $C(u)$ be
the set of all its descendants in $N$ and $N(u)$ the subgraph of $N$
supported on $C(u)$: it is still a phylogenetic network, with root $u$
and leaves labeled in the subset $C_L(u)\subseteq S$ of labels of the
leaves that are descendants of $u$.  We shall call $N(u)$ the
\emph{rooted subnetwork of $N$ generated by $u$}, and the set of
leaves $C_{L}(u)$ the \emph{cluster} of $u$.

 A \emph{clade} of a phylogenetic network $N$ is a rooted subnetwork
 of $N$ all whose nodes  are tree nodes in $N$ (and, in particular, it
 is a rooted tree).

Let $S$ be again a finite set of labels and $\mathcal{P}^+(S)$ the set of its non-empty subsets. A (\emph{rooted}) \emph{multi-labeled phylogenetic tree} (a \emph{MUL-tree}, for short) over $S$ is a rooted tree whose leaves are labeled in $\mathcal{P}^+(S)$. 
In particular, two leaves in a MUL-tree may share one or more labels.  More in general, a \emph{multi-labelled DAG} over $S$ is a DAG whose leaves are labeled in $\mathcal{P}^+(S)$. 

\subsection{Multisets and metrics}

Let $\mathcal{C}$ be a class endowed with a notion of isomorphism $\cong$; for instance, the class of all phylogenetic networks on a given set of taxa. A \emph{metric} on $\mathcal{C}$ is a mapping 
$$
d:\mathcal{C}\times \mathcal{C} \to \RR
$$
satisfying the following axioms: for every $A,B,C\in \mathcal{C}$,
\begin{enumerate}[(a)]
\item \emph{Non-negativity}: $d(A,B)\geq 0$
\item \emph{Separation}: $d(A,B)=0$ if and only if $A\cong B$
\item \emph{Symmetry}: $d(A,B)=d(B,A)$
\item \emph{Triangle inequality}: $d(A,C)\leq d(A,B)+d(B,C)$
\end{enumerate}

A \emph{finite multiset} of elements of a set $X$ is a mapping $M:X\to \NN$ such that its \emph{support}
$\{x\in X\mid M(x)\neq 0\}$ is finite. If the support of a finite multiset $M:X\to \NN$ is $\{x_1,\ldots,x_k\}$, then  this multiset can be understood as a (sort of) set  consisting of $M(x_i)$ copies of $x_i$, for every $i=1,\ldots,k$; in this context, $M(x)$ is called the \emph{multiplicity} of $x\in X$ in this multiset, and this multiplicity is 0 when $x$ does not belong to the support.

The \emph{cardinal} $|M|$ of a finite multiset $M$ of elements of $X$ is simply the sum of the multiplicities of the elements:
$$
|M|=\sum_{x\in X} M(x).
$$

The \emph{symmetric difference} of two finite multisets $M_1,M_2$ of elements of a set $X$ is the finite multiset
$$
\begin{array}{rcl}
M_1\bigtriangleup M_2 : X & \to & \NN\\ 
x & \mapsto & |M_1(x)-M_2(x)|
\end{array}
$$
Thus, if an element of $X$ has multiplicity $m_1$ in  $M_1$ and $m_2$ in $M_2$, then
it has multiplicity $|m_1-m_2|$ in $M_1\bigtriangleup M_2$.

Given a set $X$, we shall denote by $\mathcal{M}(X)$ the class of all finite multisets of elements of $X$. The mapping
$$
\begin{array}{rcl}
d: \mathcal{M}(X)\times \mathcal{M}(X)& \to & \RR\\
(M_1,M_2) & \mapsto & |M_1 \bigtriangleup M_2 |
\end{array}
$$
that associates to each pair of finite multisets the cardinal of their symmetric difference, is a metric on $\mathcal{M}(X)$, taking as notion of isomorphism the equality of multisets; this metric is called the  \emph{symmetric difference metric} on $\mathcal{M}(X)$  (see, for instance, \cite[p. 25]{deza2006dd} for the general version on a measure space). Since the condition of being a metric is not affected by the multiplication by an scalar factor, $\frac{1}{2}d$ is also a metric on $\mathcal{M}(X)$.

 We shall use several times, usually without any further notice, the following easy result.

\begin{proposition}\label{prop:metrics}
Let 
$F:\mathcal{C}\to \mathcal{M}(X)$ be a mapping such that
if $A\cong B$, then $F(A)= F(B)$. Then, the mapping
$$
\begin{array}{rcl}
d_F: \mathcal{C}\times \mathcal{C} & \to & \RR\\
(A,B) & \mapsto & \frac{1}{2}|F(A)\bigtriangleup F(B)|
\end{array}
$$
 is a metric on $\mathcal{C}$ if, and only if, 
it satisfies the following condition:
\begin{itemize}
\item If $F(A)=F(B)$, then $A\cong B$.
\end{itemize}
\end{proposition}

\begin{proof}
Notice that $d_F(A,B)=\frac{1}{2}d(F(A),F(B))$.  Then, the non-negativity, symmetry and triangle inequality axioms for $d_F$ are derived from the corresponding properties of $\frac{1}{2}d$, without any further assumption. As far as the separation axiom goes, 
if $A\cong B$, then $F(A)= F(B)$ and hence $d_F(A,B)=0$, also without any further assumption on $F$. The converse implication in the separation axiom says 
$$
|F(A)\bigtriangleup F(B)|=0\mbox{ implies } A\cong B,
$$
and since (by the separation axiom for the symmetric difference metric) $|F(A)\bigtriangleup F(B)|=0$ is equivalent to $F(A)= F(B)$, it is clear that this remaining condition is equivalent to the condition given in the statement. \qed
\end{proof}

\subsection{The $\mu$-distance}

Let $N=(V,E)$ be a phylogenetic network on $S=\{1,\dots,n\}$.  For
every node $v\in V$ and for every $i=1,\ldots,n$, let $m_i(v)$ the
number of different paths from $v$ to the leaf $i$.  The
\emph{path-multiplicity vector}, or \emph{$\mu$-vector}, for short, of
$v\in V$ is
$$
\mu(v)=(m_1(v),\dots,m_n(v)).
$$
The \emph{$\mu$-representation} of $N$ is the multiset
$$
\mu(N)=\{\mu(v)\mid v\in V\},
$$
where every vector appears with multiplicity the number of nodes having it as their $\mu$-vector.

The \emph{$\mu$-distance} between a pair of phylogenetic networks $N_1$ and $N_2$ on the same set $S$ of taxa is
$$
d_\mu(N_1,N_2)=\frac{1}{2}|\mu(N_1) \bigtriangleup \mu(N_2)|,
$$
where $\bigtriangleup $ denotes the symmetric difference of multisets.

This $\mu$-distance is known to be a metric on several well-defined classes of phylogenetic networks.
More specifically, we have the following result (and then Proposition \ref{prop:metrics} applies).

\begin{theorem}\label{thm:mu}
Let $N_1$ and $N_2$ be two phylogenetic networks on the same set $S$ of taxa. Assume that one of the following two conditions holds:
\begin{enumerate}[(a)]
\item $N_1$ and $N_2$ are both tree-child, or
\item $N_1$ and $N_2$ are both semi-binary, time consistent and tree-sibling.
\end{enumerate}
Then, $\mu(N_1)=\mu(N_2)$ implies $N_1\cong N_2$.\qed
\end{theorem}

For a proof of the case (a), see \cite{cardona.ea:07b}, and for the case (b), see
\cite{cardona.ea:sbTSTC:2008}.



\subsection{Moret-Nakhleh-Warnow-\textsl{et al}'s reduction process}

Let $N=(V,E)$ be a phylogenetic network on a set $S$ of taxa.  Two nodes in $N$ are said to be \emph{convergent} when they have
the same cluster.
The removal of convergent sets is the basis of the following  \emph{reduction
procedure}  introduced in \cite{moret.ea:2004}. 

 Let $N=(V,E)$ be a phylogenetic network on $S$. If $N$ does not contain any pair of convergent nodes (for instance, if it is a phylogenetic tree), then the reduction procedure does nothing. Otherwise:
\begin{enumerate}
\item[(0)] For every clade $T$ of $N$, with root $r_T$:
\begin{itemize}
\item Add a new node $h_T$ between $r_T$ and its  only parent.

\item Label $h_T$ with some symbol representing the clade $T$.

\item Remove $r_T$ and its descendants, so that $h_T$ becomes a leaf: we shall call it a \emph{symbolic leaf}.
\end{itemize}

After this step, the resulting multi-labeled DAG $N^*$ has two kinds of leaves: symbolic, which replace clades, and hybrid, which did not belong to any clade in $N$ (the \emph{reconstructible} phylogenetic networks considered in  \cite{moret.ea:2004} could not contain hybrid leaves, but they can be handled without any problem by the reduction procedure). 

\item[(1)] All internal nodes that are convergent in $N$ with some other node are removed from $N^*$, and all internal nodes of $N^*$ that are descendant of some removed node are also removed.

\item[(2)] For every remaining node $x$ in $N^*$ that was a parent of a
node $v$ that has been removed in (1), add a new arc from $x$ to every (hybrid or symbolic) leaf that was a descendant of $v$ in $N^*$, if such an arc does not exist yet.

The resulting DAG contains no set of convergent nodes, because
any pair of convergent nodes in it would have already been convergent in $N$.

\item[(3)] For every symbolic leaf $h_T$, unlabel it and append to it  the
corresponding clade $T$, with an arc from $h_T$ to $r_T$.

\item[(4)] Replace every node  with only one parent and one child by an arc
from its parent to its only child.

Since the DAG resulting from (2) contains no pair of convergent nodes,
it contains no node with only one child.  Therefore the only possible
nodes with only one parent and one child after step (3) are those that
were symbolic leaves with only one parent.  These are the only nodes
that have to be removed in this step.
\end{enumerate}

Notice that the effect of (3) and (4) is not exactly the replacement of each symbolic leaf by the corresponding clade: the symbolic leaf $h_T$ survives after (4) if it has more than one incoming arc, and in this case the clade $T$ is appended to $h_T$, instead of replacing it.

The output of this procedure applied to a phylogenetic
network $N$ on $S$ is a (non necessarily rooted) leaf-labeled DAG, called
the \emph{reduced version} of $N$ and denoted by $R(N)$.  

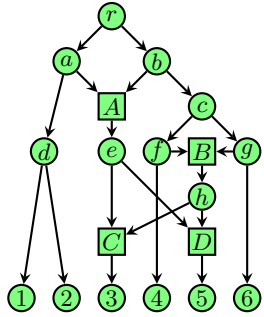
\begin{figure}[htb]
\begin{center}
\begin{tikzpicture}[thick,>=stealth,scale=0.6]
\draw (1,0) node[tre] (r) {}; \etq r
\draw (0,-1) node[tre] (a) {}; \etq a 
\draw (2,-1) node[tre] (b) {}; \etq b 
\draw (1,-2) node[hyb] (A) {}; \etq A 
\draw (3,-2) node[tre] (c) {}; \etq c 

\draw (1,-3) node[tre] (e) {}; \etq e
\draw (2,-3) node[tre] (f) {}; \etq f 
\draw (3,-3) node[hyb] (B) {}; \etq B 
\draw (4,-3) node[tre] (g) {}; \etq g
\draw (3,-4) node[tre] (h) {}; \etq h 
\draw (1,-5) node[hyb] (C) {}; \etq C
\draw (3,-5) node[hyb] (D) {}; \etq D 
\draw (-0.5,-3) node[tre] (d) {}; \etq d 
\draw (-1,-6.25) node[tre] (1) {}; \etq 1
\draw (0,-6.25) node[tre] (2) {}; \etq 2
\draw (1,-6.25) node[tre] (3) {}; \etq 3
\draw (2,-6.25) node[tre] (4) {}; \etq 4
\draw (3,-6.25) node[tre] (5) {}; \etq 5
\draw (4,-6.25) node[tre] (6) {}; \etq 6
\draw [->](r)--(a);
\draw [->](r)--(b);
\draw [->](a)--(d);
\draw [->](d)--(1);
\draw [->](d)--(2);
\draw [->](a)--(A);
\draw [->](b)--(A);
\draw [->](A)--(e);
\draw [->](b)--(c);
\draw [->](c)--(f);
\draw [->](c)--(g);
\draw [->](e)--(C);
\draw [->](e)--(D);
\draw [->](f)--(4);
\draw [->](f)--(B);
\draw [->](g)--(B);
\draw [->](g)--(6);
\draw [->](B)--(h);
\draw [->](h)--(C);
\draw [->](h)--(D);
\draw [->](C)--(3);
\draw [->](D)--(5);
\end{tikzpicture}
\end{center}
\caption{\label{fig:ex-red1} The phylogenetic network $N$ in Example~\ref{ex:red}.}
\end{figure}

\begin{example}\label{ex:red}
Let us compute the reduced version of  the phylogenetic network $N$ represented in
Fig.~\ref{fig:ex-red1}. The subtree rooted at $d$, with leaves 1 and 2, is a clade, and each one of the other leaves forms a clade by itself. The graph $N^*$ obtained after step (0) is depicted in 
Fig.~\ref{fig:ex-red2}.(0), where the symbolic leaves that replace clades are represented by dashed circles.

The maximal  sets of convergent nodes in
$N$ are
$$
\{b,c\}, \{A,B,e,h\}, \{C,2\}, \{D,4\}.
$$
So, in step (1) we remove  the  nodes $b,c,e,h,A,B,C,D$, as well as all
intermediate nodes in paths from them to symbolic leaves: this also
removes the nodes $f,g$.  So, the only internal nodes that remain
after after step (1) are $r$ and $a$.  This yields the graph depicted  in Fig.~\ref{fig:ex-red2}.(1).

In step (2),  we add new arcs from $r$ and $a$ to the
symbolic leaves that were descendant of removed descendants of them. This yields the graph  
 in Fig.~\ref{fig:ex-red2}.(2). 

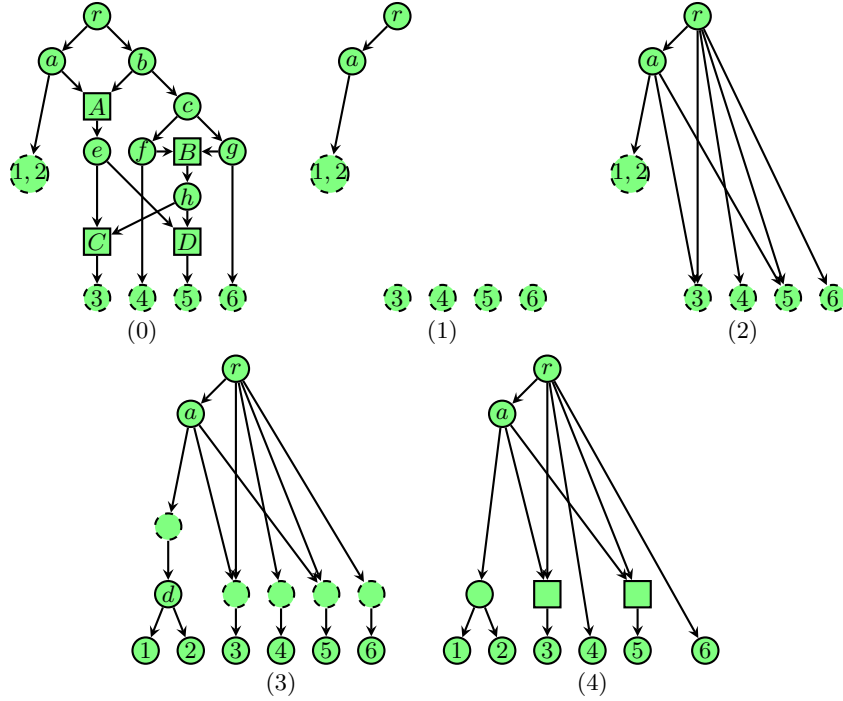
\begin{figure}[htb]
\begin{center}
\begin{tikzpicture}[thick,>=stealth,scale=0.6]
\draw (1,0) node[tre] (r) {}; \etq r
\draw (0,-1) node[tre] (a) {}; \etq a 
\draw (2,-1) node[tre] (b) {}; \etq b 
\draw (1,-2) node[hyb] (A) {}; \etq A 
\draw (3,-2) node[tre] (c) {}; \etq c 
\draw (1,-3) node[tre] (e) {}; \etq e
\draw (2,-3) node[tre] (f) {}; \etq f 
\draw (3,-3) node[hyb] (B) {}; \etq B 
\draw (4,-3) node[tre] (g) {}; \etq g
\draw (3,-4) node[tre] (h) {}; \etq h 
\draw (1,-5) node[hyb] (C) {}; \etq C
\draw (3,-5) node[hyb] (D) {}; \etq D 
\draw (-0.5,-3.5) node[circle,fill=green!50,draw,minimum size=5mm,style=dashed] (1s) {}; 
\draw (1s) node {\footnotesize $1,2$};
\draw (1,-6.25) node[circle,fill=green!50,draw,minimum size=3.5mm,style=dashed] (2s) {};
\draw (2s) node {\footnotesize $3$};
\draw (2,-6.25) node[circle,fill=green!50,draw,minimum size=3.5mm,style=dashed] (3s) {}; 
\draw (3s) node {\footnotesize $4$};
\draw (3,-6.25) node[circle,fill=green!50,draw,minimum size=3.5mm,style=dashed] (4s) {}; 
\draw (4s) node {\footnotesize $5$};
\draw (4,-6.25) node[circle,fill=green!50,draw,minimum size=3.5mm,style=dashed] (5s) {}; 
\draw (5s) node {\footnotesize $6$};
\draw [->](r)--(a);
\draw [->](r)--(b);
\draw [->](a)--(1s);
\draw [->](a)--(A);
\draw [->](b)--(A);
\draw [->](A)--(e);
\draw [->](b)--(c);
\draw [->](c)--(f);
\draw [->](c)--(g);
\draw [->](e)--(C);
\draw [->](e)--(D);
\draw [->](f)--(3s);
\draw [->](f)--(B);
\draw [->](g)--(B);
\draw [->](g)--(5s);
\draw [->](B)--(h);
\draw [->](h)--(C);
\draw [->](h)--(D);
\draw [->](C)--(2s);
\draw [->](D)--(4s);
\draw (2,-7) node {(0)}; 
\end{tikzpicture}
\qquad
\begin{tikzpicture}[thick,>=stealth,scale=0.6]
\draw (1,0) node[tre] (r) {}; \etq r
\draw (0,-1) node[tre] (a) {}; \etq a 
\draw (-0.5,-3.5) node[circle,fill=green!50,draw,minimum size=5mm,style=dashed] (1s) {}; 
\draw (1s) node {\footnotesize $1,2$};
\draw (1,-6.25) node[circle,fill=green!50,draw,minimum size=3.5mm,style=dashed] (2s) {};
\draw (2s) node {\footnotesize $3$};
\draw (2,-6.25) node[circle,fill=green!50,draw,minimum size=3.5mm,style=dashed] (3s) {}; 
\draw (3s) node {\footnotesize $4$};
\draw (3,-6.25) node[circle,fill=green!50,draw,minimum size=3.5mm,style=dashed] (4s) {}; 
\draw (4s) node {\footnotesize $5$};
\draw (4,-6.25) node[circle,fill=green!50,draw,minimum size=3.5mm,style=dashed] (5s) {}; 
\draw (5s) node {\footnotesize $6$};
\draw [->](r)--(a);
\draw [->](a)--(1s);
\draw (2,-7) node {(1)}; 
\end{tikzpicture}
\qquad
\begin{tikzpicture}[thick,>=stealth,scale=0.6]
\draw (1,0) node[tre] (r) {}; \etq r
\draw (0,-1) node[tre] (a) {}; \etq a 
\draw (-0.5,-3.5) node[circle,fill=green!50,draw,minimum size=5mm,style=dashed] (1s) {}; 
\draw (1s) node {\footnotesize $1,2$};
\draw (1,-6.25) node[circle,fill=green!50,draw,minimum size=3.5mm,style=dashed] (2s) {};
\draw (2s) node {\footnotesize $3$};
\draw (2,-6.25) node[circle,fill=green!50,draw,minimum size=3.5mm,style=dashed] (3s) {}; 
\draw (3s) node {\footnotesize $4$};
\draw (3,-6.25) node[circle,fill=green!50,draw,minimum size=3.5mm,style=dashed] (4s) {}; 
\draw (4s) node {\footnotesize $5$};
\draw (4,-6.25) node[circle,fill=green!50,draw,minimum size=3.5mm,style=dashed] (5s) {}; 
\draw (5s) node {\footnotesize $6$};
\draw [->](r)--(a);
\draw [->](r)--(2s);
\draw [->](r)--(3s);
\draw [->](r)--(4s);
\draw [->](r)--(5s);
\draw [->](a)--(1s);
\draw [->](a)--(2s);
\draw [->](a)--(4s);
\draw (2,-7) node {(2)}; 
\end{tikzpicture}\\
\begin{tikzpicture}[thick,>=stealth,scale=0.6]
\draw (1,0) node[tre] (r) {}; \etq r
\draw (0,-1) node[tre] (a) {}; \etq a 
\draw (-0.5,-3.5) node[circle,fill=green!50,draw,minimum size=3.5mm,style=dashed] (1s) {}; 
\draw (1,-5) node[circle,fill=green!50,draw,minimum size=3.5mm,style=dashed] (2s) {};
\draw (2,-5) node[circle,fill=green!50,draw,minimum size=3.5mm,style=dashed] (3s) {}; 
\draw (3,-5) node[circle,fill=green!50,draw,minimum size=3.5mm,style=dashed] (4s) {}; 
\draw (4,-5) node[circle,fill=green!50,draw,minimum size=3.5mm,style=dashed] (5s) {}; 
\draw (-0.5,-5) node[tre] (d) {}; \etq d 
\draw (-1,-6.25) node[tre] (1) {}; \etq 1
\draw (0,-6.25) node[tre] (2) {}; \etq 2
\draw (1,-6.25) node[tre] (3) {}; \etq 3
\draw (2,-6.25) node[tre] (4) {}; \etq 4
\draw (3,-6.25) node[tre] (5) {}; \etq 5
\draw (4,-6.25) node[tre] (6) {}; \etq 6
\draw [->](r)--(a);
\draw [->](r)--(2s);
\draw [->](r)--(3s);
\draw [->](r)--(4s);
\draw [->](r)--(5s);
\draw [->](a)--(1s);
\draw [->](a)--(2s);
\draw [->](a)--(4s);
\draw [->](1s)--(d);
\draw [->](2s)--(3);
\draw [->](3s)--(4);
\draw [->](4s)--(5);
\draw [->](5s)--(6);
\draw [->](d)--(1);
\draw [->](d)--(2);
\draw (2,-7) node {(3)}; 
\end{tikzpicture}
\qquad
\begin{tikzpicture}[thick,>=stealth,scale=0.6]
\draw (1,0) node[tre] (r) {}; \etq r
\draw (0,-1) node[tre] (a) {}; \etq a 
\draw (1,-5) node[hyb] (2s) {};
\draw (3,-5) node[hyb] (4s) {}; 
\draw (-0.5,-5) node[tre] (d) {}; 
\draw (-1,-6.25) node[tre] (1) {}; \etq 1
\draw (0,-6.25) node[tre] (2) {}; \etq 2
\draw (1,-6.25) node[tre] (3) {}; \etq 3
\draw (2,-6.25) node[tre] (4) {}; \etq 4
\draw (3,-6.25) node[tre] (5) {}; \etq 5
\draw (4.5,-6.25) node[tre] (6) {}; \etq 6
\draw [->](r)--(a);
\draw [->](r)--(2s);
\draw [->](r)--(4s);
\draw [->](a)--(d);
\draw [->](a)--(2s);
\draw [->](a)--(4s);
\draw [->](2s)--(3);
\draw [->](r)--(4);
\draw [->](4s)--(5);
\draw [->](r)--(6);
\draw [->](d)--(1);
\draw [->](d)--(2);
\draw (2,-7) node {(4)}; 
\end{tikzpicture}

\end{center}
\caption{\label{fig:ex-red2} The resulting DAGs after the different steps in the reduction process applied to $N$.}
\end{figure}

In step (3), we append again to each symbolic leaf the clade it represented, and we unlabel the symbolic leaves: see 
 Fig.~\ref{fig:ex-red2}.(3).

Finally, in step (4) the parents of the node $d$ and of the leaves 4 and 6 are removed and replaced by arcs $(a,d)$, $(r,4)$, and $(r,6)$, respectively.
The parents of leaves 3 and 5 remain, and they are hybrid in the
resulting reduced network $R(N)$, which is given in
 Fig.~\ref{fig:ex-red2}.(4).
\end{example}

A phylogenetic network
$N$ is \emph{reduced} when $R(N)=N$. From the given description of reduction procedure, it is easy to deduce that a
phylogenetic network is reduced if, and only if, every  pair of convergent nodes in it consists of a hybrid node of out-degree 1 and with all its proper descendants  of tree type (thus forming a clade), and its only child. In particular, if we impose that all hybrid nodes in a phylogenetic network have out-degree 1, as it is  done for instance in reconstructible networks in the sense of  \cite{moret.ea:2004}, then a reduced network cannot contain any hybrid node that is a proper descendant of another hybrid node.

Two networks $N_{1}$ and $N_{2}$ are said to be
\emph{indistinguishable} when they have isomorphic reduced versions,
that is, when $R(N_{1})\cong R(N_{2})$.  Moret, Nakhleh, Warnow, et al
argue in \cite[p.  19]{moret.ea:2004} that for reconstructible
phylogenetic networks this notion of indistinguishability (isomorphism
after simplification) is more suitable than the existence of an
isomorphism between the original networks.

\section{Nakhleh's `metric' $m$}

In this section we describe the dissimilarity measure $m$ introduced by Nakhleh  in  \cite{nakhleh:2007}. 
After recalling Nakhleh's definition, we provide an alternative definition,  as the cardinal of the symmetric difference of certain representations of the networks, which allows simpler proofs of the new results reported in this paper.

Nakhleh begins by defining the following equivalence of nodes in pairs of phylogenetic networks.

\begin{definition}
Let $N_1=(V_1,E_1)$ and $N_2=(V_2,E_2)$ be a pair of phylogenetic networks (not necessarily
different). Two nodes $u\in V_1$ and $v\in V_2$
are \emph{equivalent}, in symbols $u\equiv v$, when:
\begin{itemize}
\item $u$ and $v$ are both leaves labeled with the same taxon, or

\item for some $k\geq 1$, node $u$ has  exactly $k$ children $u_1,\ldots,u_k$,
node $v$ has exactly $k$ children $v_1,\ldots, v_k$, and $u_i\equiv v_i$ for
every $i=1,\ldots,k$.
\end{itemize}
For every node $v$ in a phylogenetic network, let $\kappa(v)$ be the number of nodes in this network that are equivalent to it.
\end{definition}

Then, he defines his dissimilarity
measure by comparing the cardinals of equivalence classes of nodes in pairs of phylogenetic networks.

\begin{definition}
Let $N_1$ and $N_2$ be a pair of phylogenetic networks on the same set $S$ of taxa, and let $U(N_1)$ and $U(N_2)$ be maximal sets of non-equivalent nodes in them. For every $v_1\in U(N_1)$, let 
$$
\delta(v_1)=\left\{
\begin{array}{ll}
\kappa(v_1) & \mbox{ if no node in $U(N_2)$ is equivalent to $v_1$}\\
\max\{0, \kappa(v_1)-\kappa(v_1')\} & \mbox{ if $v_1'\in U(N_2)$ is equivalent to $v_1$}
\end{array}
\right.
$$
The value $\delta(v_2)$, for every $v_2\in U(N_2)$, is defined in a similar way.

Then, let
$$
m(N_1,N_2)=\frac{1}{2}\Big(\sum_{v_1\in U(N_1)} \delta(v_1)+\sum_{v_2\in U(N_2)} \delta(v_2)\Big).
$$
\end{definition}

To introduce our version of this metric, we define first a nested labeling of the nodes of a phylogenetic network.

\begin{definition}
Let $N=(V,E)$ be a phylogenetic network on a set $S$ of taxa. The \emph{nested label} $\ell(v)$ of a
node $v$ of $N$ is defined by induction on $h(v)$ as follows:

\begin{itemize}
\item If $h(v)=0$, that is, if $v$ is a leaf, then $\ell(v)$ is the singleton consisting of its label.

\item If $h(v)=m>0$, then all its children $v_1,\ldots,v_k$ have
height smaller than $m$, and hence they have been already labeled:
then, $\ell(v)$ is the multiset of their nested labels,
$$
\ell(v)=\{\ell(v_1),\ldots,\ell(v_k)\}.
$$
\end{itemize}
\end{definition}

Notice that the nested label of a node is, in general, a nested
multiset (a multiset of multisets of multisets of\ldots), hence its
name.  Moreover, the height of a node $u$ is the highest level of
nesting of a leaf in $\ell(u)$ minus 1, and the cluster of $u$ consists of the taxa appearing in $\ell(u)$.

\begin{example}
Table \ref{table:ex-red1} gives the nested labels of the nodes of
the phylogenetic network depicted in Fig.~\ref{fig:ex-red1}, sorted by their height. 
\end{example}
\begin{center}
 \begin{table}[htdp]
\caption{Nested labels of the nodes of the phylogenetic network $N$ in Fig.~\ref{fig:ex-red1}.}
{\footnotesize\begin{tabular}{|c|l|}
\hline
1 & $\{1\}$ \\
\hline
2 & $\{2\}$ \\
\hline
3 & $\{3\}$ \\
\hline
4 & $\{4\}$ \\
\hline
5 & $\{5\}$ \\
\hline
6 & $\{6\}$ \\
\hline
$d$ & $\{\{1\},\{2\}\}$\\
\hline
$C$ & $\{\{3\}\}$ \\
\hline
$D$ & $\{\{5\}\}$ \\
\hline
$e$ & $\{\{\{3\}\},\{\{5\}\}\}$ \\
\hline
$h$ & $\{\{\{3\}\},\{\{5\}\}\}$ \\
\hline
$A$ & $\{\{\{\{3\}\},\{\{5\}\}\}\}$ \\
\hline
$B$ & $\{\{\{\{3\}\},\{\{5\}\}\}\}$ \\
\hline
$f$ & $\{\{4\},\{\{\{\{3\}\},\{\{5\}\}\}\}\}$ \\
\hline
$g$ & $\{\{\{\{\{3\}\},\{\{5\}\}\}\},\{6\}\}$ \\
\hline
$c$ & $\{\{\{\{\{\{3\}\},\{\{5\}\}\}\},\{6\}\},\{\{4\},\{\{\{\{3\}\},\{\{5\}\}\}\}\}\}$ \\
\hline
 $a$ & $\{\{\{1\},\{2\}\},\{\{\{\{3\}\},\{\{5\}\}\}\}\}$\\
\hline
$b$ & $\{\{\{\{\{3\}\},\{\{5\}\}\}\},\{\{\{\{\{\{3\}\},\{\{5\}\}\}\},\{6\}\},\{\{4\},\{\{\{\{3\}\},\{\{5\}\}\}\}\}\}\}$ \\
\hline
$r$  & $\{\{\{\{1\},\{2\}\},\{\{\{\{3\}\},\{\{5\}\}\}\}\},$\\
 & \qquad\qquad\qquad\qquad $\{\{\{\{\{3\}\},\{\{5\}\}\}\},\{\{\{\{\{\{3\}\},\{\{5\}\}\}\},\{6\}\},\{\{4\},\{\{\{\{3\}\},\{\{5\}\}\}\}\}\}\}\}$ \\
\hline
\end{tabular}
}
\label{table:ex-red1}
\end{table}%
\end{center}

We shall say that a nested label $\ell(v)$ is \emph{contained} in a
nested label $\ell(u)$, in symbols $\ell(v)\contained \ell(u)$, when $\ell(v)$ is the nested label of a
descendant of $u$.  Notice that the fact that $\ell(v)$ is contained in
$\ell(u)$, does not imply that $v$ is a descendant of $u$:
several instances of this fact can be detected in the network represented in
Fig.~\ref{fig:ex-red1}. Notice moreover that $\ell(v)\in \ell(u)$ if, and only if,
$\ell(v)$ is the nested label of a child of $u$.

Nakhleh's equivalence relation is easily characterized in terms of nested labels.

\begin{proposition}
Let $N_1=(V_1,E_1)$ and $N_2=(V_2,E_2)$ be a pair of phylogenetic networks (not necessarily
different) labeled in a set $S$.  For every $u\in V_1$ and $v\in V_2$,
$u\equiv v$ if, and only if, $\ell(u)=\ell(v)$.
\end{proposition}

\begin{proof}
We prove the equivalence by induction on the height of one of the
nodes, say $u$.

If $h(u)=0$, then it is a leaf, and $\ell(u)$ is the one-element set
consisting of its label.  Thus, in this case, $u\equiv v$ if, and only
if, $v$ is the leaf of $N_2$ with the same label as $u$, and
$\ell(u)=\ell(v)$ if, and only if, $v$ is the leaf of  $N_2$ with the
same label as $u$, too.

Consider now the case when $h(u)=m>0$ and assume that the thesis holds
for all nodes $u'\in V_1$ of height smaller than $m$.  Let
$u_1,\ldots, u_k$ be the children of $u$.  Then:
\begin{itemize}
\item $u\equiv v$ if and only if $v$ has  exactly $k$ children and they can be
ordered $v_1,\ldots,v_k$ in such a way that $u_i\equiv v_i$ for every
$i=1,\ldots,k$.

\item $\ell(u)=\ell(v)$ if and only if $v$ has exactly  $k$ children and the
multiset of their nested labels is equal to the multiset of nested
labels of $u_1,\ldots,u_k$, which means that $v$'s children can be
ordered $v_1,\ldots,v_k$ in such a way that $\ell(u_i)=\ell(v_i)$ for
every $i=1,\ldots,k$.
\end{itemize}
Since, by induction, the children of $u$ satisfy the thesis, it is
clear that $u\equiv v$ is equivalent to $\ell(u)=\ell(v)$.\qed
\end{proof}

Thus, we can rewrite Nakhleh's dissimilarity measure in terms of nested labels.

\begin{definition}
For every $S$-DAG $N$, the \emph{nested labels representation} of $N$ is the multiset $\Upsilon(N)$ of nested labels of its nodes (where each nested label appears with
multiplicity the number of nodes having it).
\end{definition}

\begin{proposition}\label{prop:m-nl}
For every pair $N_1,N_2$ of phylogenetic networks over the same set $S$ of taxa,
$$
m(N_1,N_2)=\frac{1}{2}|\Upsilon(N_1) \bigtriangleup \Upsilon(N_2)|,
$$
where $\bigtriangleup $ denotes the symmetric difference of multisets.
\end{proposition}

\begin{proof}
Let $N_1,N_2$ be a pair of phylogenetic networks over the same set $S$ of taxa, and $U(N_1), U(N_2)$ maximal sets of non-equivalent nodes in them.
Since the equivalence of nodes is synonymous of having the same nested labels, it is clear that, for every $i=1,2$, $\Upsilon(N_i)$ is the multiset consisting of $\kappa(v)$ copies of $\ell(v)$, for each $v\in U(N_i)$. Then:
\begin{itemize}
\item If $v_1\in U(N_1)$ is not equivalent to any node in $U(N_2)$, then $\ell(v_1)$ contributes $\kappa(v_1)=\delta(v_1)$ to $|\Upsilon(N_1) \bigtriangleup \Upsilon(N_2)|$.

\item If $v_2\in U(N_2)$ is not equivalent to any node in $U(N_1)$, then $\ell(v_2)$ contributes $\kappa(v_2)=\delta(v_2)$ to $|\Upsilon(N_1) \bigtriangleup \Upsilon(N_2)|$.

\item If $v_1\in U(N_1)$ is equivalent to $v_2\in U(N_2)$, then $\ell(v_1)=\ell(v_2)$ contributes 
$$
|\kappa(v_1)-\kappa(v_2)|=\max\{0, \kappa(v_1)-\kappa(v_2)\} +\max\{0, \kappa(v_2)-\kappa(v_1)\} 
=\delta(v_1)+\delta(v_2)
$$
 to $|\Upsilon(N_1) \bigtriangleup \Upsilon(N_2)|$.
\end{itemize}
This implies that
$$
|\Upsilon(N_1) \bigtriangleup \Upsilon(N_2)|=
\sum_{v_1\in U(N_1)} \delta(v_1)+\sum_{v_2\in U(N_2)} \delta(v_2),
$$
from where the equality $m(N_1,N_2)=\frac{1}{2}|\Upsilon(N_1) \bigtriangleup \Upsilon(N_2)|$ follows. \qed 
\end{proof}

In the rest of this paper, we shall  use the definition of  $m$ provided by the last proposition.

The value $m(N_1,N_2)$ can be computed in time polynomial in
the sizes of the networks $N_1,N_2$  by performing a simultaneous bottom-up traversal of the two
networks \cite{valiente:2002,valiente:2007}

\section{The separation power of $m$}

\subsection{$m$ separates distinguishable networks}
Nakhleh  proved  in  \cite[Thm.~2]{nakhleh:2007}  the following result. 

\begin{proposition}
\label{prop:nak}
Let $R(N_1)$ and $R(N_2)$ be the reduced versions of two phylogenetic
networks on the same set $S$ of taxa.  Then, $m(R(N_1),R(N_2))=0$ if, and only if, $R(N_1)\cong
R(N_2)$. \qed
\end{proposition}

In this subsection we extend this result by showing that   $m$  separates
phylogenetic networks that are distinguishable up to reduction; a sketch of this proof can be found in our previous preprint \cite{cardona.ea:08-2metrics}.  We would like to
recall here that this was the (unaccomplished: see
\cite{cardona.ea:07a}) goal of the error metric defined in
\cite{moret.ea:2004}.

\begin{theorem}
Let $N_1$ and $N_2$ be two phylogenetic networks on the set $S$ of
taxa. If
$m(N_1,N_2)=0$, then $N_1$ and $N_2$ are indistinguishable.
\end{theorem}

\begin{proof}
 Let $N_1=(V_1,E_1)$ and $N_2=(V_2,E_2)$ be two phylogenetic
networks such that $\Upsilon(N_1)=\Upsilon(N_2)$.  We shall prove that
the reduction process of both networks modifies exactly in the same
way their nested labels representations, and thus the reduced
versions $R(N_1)$ and $R(N_2)$ are also such that $\Upsilon (R(N_1))= \Upsilon (R(N_2))$.  Then, by Proposition \ref{prop:nak}, the latter are
isomorphic.

To begin with, notice that two nodes are convergent when the sets of
taxa appearing in their nested labels are the same (without
taking into account nesting levels or multiplicities).  In particular,
$N_1$ and $N_2$ have the same sets of nested labels of convergent
nodes.

Step (0) in the reduction process consists of replacing every clade by
a symbolic leaf.  This corresponds to remove the
nested labels of the nodes belonging to clades (except their roots)
and to replace, in all remaining nested labels, each nested label of a
root of a clade by the label of the corresponding symbolic leaf. We 
must prove now that we can decide from the nested labels representations alone which 
are the nested labels of nodes of clades and of roots of clades. 

Since  the clades of a phylogenetic network are subtrees, a node
belonging to a clade is only equivalent to itself (if $v$ is a node of a clade and $\ell(u)=\ell(v)$, then $C_L(u)=C_L(v)$, but in this case, since $v$ is the least common ancestor of $C_L(v)$ in the clade it belongs,
$v$ must be a descendant of $u$, and since $u$ and $v$ have the same height ---because they have the same nested label-- they must be the same node). In particular, a node of a clade does not share its nested label with any other node.

Then, 
the nested labels of nodes $v\in V_i$  belonging to some clade of $N_i$ ($i=1,2$) are characterized by
the following two properties: 
$\ell(v)$ and each one of  the nested labels contained in it appear with multiplicity 1 in $\Upsilon(N_1)=\Upsilon(N_2)$ (and in particular $v$ and its descendants are characterized by their nested labels);
and  $\ell(v)$ and each one of  the nested labels contained in it belong at most to one nested label
(this means that $v$ and its descendants are tree nodes, and in particular that the rooted subnetwork generated by $v$ is a tree consisting only of tree nodes from $N_i$).
And therefore the roots of clades of $N_i$ are the nodes $v$ with nested label $\ell(v)$ maximal with these properties, and the nodes of
the clade rooted at $v$ are those nodes with nested labels contained in
$\ell(v)$.  This shows that the nested
labels of roots of clades and the nested labels of nodes belonging to
clades in $N_{1}$ are the same as in $N_{2}$.

So, we remove the same nested labels in $N_{1}$ and $N_{2}$ and we 
replace the same nested labels by symbolic leaves. As a consequence, 
the networks resulting after this step have the same nested labels.

In step (1), all internal nodes that are convergent with some other node are
removed, and all nodes other than (symbolic or hybrid) leaves that are
descendant of some removed node are also removed.  So, in this step we
remove the nested labels other than singletons of convergent nodes, and the nested labels
other than singletons that are contained in a nested label of some
convergent node (notice that if $\ell(v)$ is not a singleton and it is
contained in $\ell(u)$ and $u$ is convergent, then either $v$ is a
descendant of $u$, and then it has to be removed, or it is equivalent
to a descendant of $u$, and then it is convergent with this
descendant and it has to be removed, too).  This shows that the nested
labels of the nodes removed in both networks are the same, and hence that  the
nested labels of the nodes that remain in both networks are also the
same.

In step (2), the paths from the remaining nodes to the labels are
restored.  It means to replace in each
remaining nested label $\ell(x)$, each maximal  
nested label $\ell(v)\contained \ell(x)$ of a removed node $v$  by the
singletons $\{s_1\},\{s_2\},\ldots,\{s_p\}$ of the leaves
appearing in $\ell(v)$.  Again, this operation only depends on the
nested labels, and therefore after this step the resulting DAGs have
the same nested labels representations.

In step (3), clades are restored. This is simply done by replacing in
the nested labels each symbolic leaf $s$ by the nested label of the
root of the clade it replaced, between brackets (because we append it
to the node corresponding to the symbolic leaf).  Since the same
clades were removed in both networks and replaced by the same symbolic leaves, after this step the resulting
DAGs still have the same nested labels representations.

Finally, in step (4), the nodes with only one parent and only one
child are removed.  This corresponds to remove nested labels of the
form $\{\{\ldots\}\}$  that are children of only one parent (that
is, that belong to only one nested
label), and replacing them in the nested labels containing them by the corresponding nested label $\{\dots\}$ without the outer brackets.
This shows that the same nested labels are removed in both DAGs and that the remaining nested labels are modified in exactly the same way.

So, at the end of this procedure, the resulting DAGs $R(N_1)$ and
$R(N_2)$ have the same nested labels representations.  By Proposition
\ref{prop:nak}, this implies that $R(N_1)$ and $R(N_2)$ are
isomorphic. \qed
\end{proof}

The converse implication is, of course false: since the reduction
process may remove parts with different topologies that yield
differences in the nested labels representations, two phylogenetic
networks with isomorphic reduced versions may have different nested labels representations.

\subsection{$m$ refines the $\mu$-distance}

As a direct consequence of Proposition \ref{prop:nak}, Nakhleh deduced that
$m$ satisfies the separation axiom of metrics on the class of all reduced 
phylogenetic networks on the same set $S$ of taxa. In this subsection we show two other independent classes of phylogenetic networks where $m$ satisfies this axiom. The key observation in our proofs is that $m$ refines the $\mu$-distance, in the sense of Proposition \ref{cor:2} below.

\begin{lemma}\label{lem:ell-mu}
Let $v$ be a node in a phylogenetic network $N$ on a set $S$ of taxa. For every $i\in S$,
$m_i(v)$ is the number of times the label $i$ appears in $\ell(v)$.
\end{lemma}

\begin{proof}
We prove it by induction on $h(v)$. If $h(v)=0$, then $v$ is a leaf, and therefore
$m_i(v)=1$ if $\ell(v)=\{i\}$ and $m_i(v)=0$ if $\ell(v)=\{j\}$, for some $j\in S\setminus\{i\}$.

Assume now that the statements is true for all nodes of height at most $m-1$, and let $v$ be a node of height $m$. Let $v_1,\ldots,v_k$ be the children of $v$, all of them of height lower than $m$. Then, on the one hand, $m_i(v)=m_i(v_1)+\cdots+m_i(v_k)$ by \cite[Lem.~4]{cardona.ea:07b}, and, on the other hand, since $\ell(v)=\{\ell(v_1),\ldots,\ell(v_k)\}$, it is clear that the number of times the label $i$ appears in $\ell(v)$ is equal to the sum of the numbers of times it appears in the nested labels $\ell(v_1),\ldots,\ell(v_k)$, which is equal, by the induction hypothesis, to $m_i(v_1)+\cdots+m_i(v_k)$. \qed
\end{proof}

\begin{corollary}
Let $N_1=(V_1,E_1)$ and $N_2=(V_2,E_2)$ be phylogenetic networks on the same set $S$ of taxa. For every $v_1\in V_1$ and $v_2\in V_2$, if $\ell(v_1)=\ell(v_2)$, then $\mu(v_1)=\mu(v_2)$.\qed
\end{corollary}

\begin{proposition}\label{cor:2}
Let $N_1=(V_1,E_1)$ and $N_2=(V_2,E_2)$ be two phylogenetic networks on the same set $S$ of taxa. Then, 
$m(N_1,N_2)\geq d_{\mu}(N_1,N_2)$.
\end{proposition}

\begin{proof}
Let us rename the nodes of $N_1$ and $N_2$ as
$$
V_1=\{v_1,v_2,\ldots,v_l,v_{l+1},\ldots,v_m,\ldots,v_s\},\
V_2=\{w_1,w_2,\ldots,w_l,w_{l+1},\ldots,w_m,\ldots,w_t\},
$$
with $|V_1|=s$ and $|V_2|=t$ and $l\leq m\leq s,t$,  in such a way that:
\begin{itemize}
\item for every $i=1,\ldots,l$, $\ell(v_i)=\ell(w_i)$ (and hence, by the last corollary, $\mu(v_i)=\mu(w_i)$), while, for every $j=l+1,\ldots, s$ and $k=l+1,\ldots,t$,
$\ell(v_j)\neq \ell(w_k)$;

\item for every $i=l+1,\ldots,m$, $\mu(v_i)=\mu(w_i)$, while, for every $j=m+1,\ldots, s$ and $k=m+1,\ldots,t$,
$\mu(v_j)\neq \mu(w_k)$.
\end{itemize}
Therefore
$$
|\Upsilon(N_1) \bigtriangleup \Upsilon(N_2)|=(s-l)+(t-l)\geq (s-m)+(t-m)=
|\mu(N_1) \bigtriangleup \mu(N_2)|,
$$
as we claimed. \qed
\end{proof}

%


\begin{corollary}
If $d_\mu$ satisfies the separation axiom on a class of phylogenetic networks, $m$ also satisfies it. 
\end{corollary}

\begin{proof}
Let $\mathcal{N}$ be a class of phylogenetic networks such that
$d_{\mu}(N_1,N_2)=0$ implies $N_1\cong N_2$ for every $N_1,N_2\in \mathcal{N}$.
Let now $N_1,N_2\in\mathcal{N}$ be such that $m(N_1,N_2)=0$. Since
$m(N_1,N_2)\geq d_{\mu}(N_1,N_2)\geq 0$, we conclude that
$d_{\mu}(N_1,N_2)=0$ and hence, by assumption, $N_1\cong N_2$. \qed
\end{proof}

Combining this result with Theorem \ref{thm:mu} we obtain the following result.

\begin{corollary}\label{cor:m-sep}
Let $N_1$ and $N_2$ be two phylogenetic networks on the same set $S$ of taxa. Assume that one of the following two conditions holds:
\begin{enumerate}[(a)]
\item $N_1$ and $N_2$ are both tree-child, or
\item $N_1$ and $N_2$ are both semi-binary, time consistent and tree-sibling.
\end{enumerate}
Then, $\Upsilon(N_1)=\Upsilon(N_2)$ implies  $N_1\cong N_2$.\qed
\end{corollary}

In particular, by Proposition \ref{prop:metrics},
 $m$ is a metric on the classes of all tree-child and of all 
semi-binary, time consistent tree-sibling phylogenetic networks.

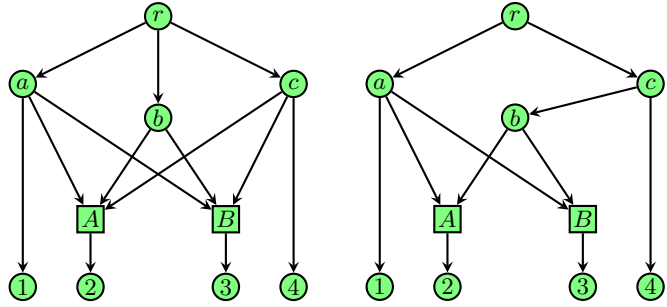
\begin{figure}[htb]
\begin{center}
\begin{tikzpicture}[thick,>=stealth,scale=0.9]
\draw (2,1) node[tre] (r) {}; \etq r 
\draw (0,0) node[tre] (a) {}; \etq a 
\draw (2,-0.5) node[tre] (b) {}; \etq b 
\draw (4,0) node[tre] (c) {}; \etq c 
\draw (1,-2) node[hyb] (A) {}; \etq A 
\draw (3,-2) node[hyb] (B) {}; \etq B 
\draw (0,-3) node[tre] (1) {}; \etq 1
\draw (1,-3) node[tre] (2) {}; \etq 2
\draw (3,-3) node[tre] (3) {}; \etq 3
\draw (4,-3) node[tre] (4) {}; \etq 4
\draw [->](r)--(a);
\draw [->](r)--(b);
\draw [->](r)--(c);
\draw [->](a)--(1);
\draw [->](a)--(A);
\draw [->](a)--(B);
\draw [->](b)--(A);
\draw [->](b)--(B);
\draw [->](c)--(A);
\draw [->](c)--(B);
\draw [->](c)--(4);
\draw [->](A)--(2);
\draw [->](B)--(3);
\end{tikzpicture}
\qquad
\begin{tikzpicture}[thick,>=stealth,scale=0.9]
\draw (2,0) node[tre] (r) {}; \etq r 
\draw (0,-1) node[tre] (a) {}; \etq a 
\draw (2,-1.5) node[tre] (b) {}; \etq b 
\draw (4,-1) node[tre] (c) {}; \etq c 
\draw (1,-3) node[hyb] (A) {}; \etq A 
\draw (3,-3) node[hyb] (B) {}; \etq B 
\draw (0,-4) node[tre] (1) {}; \etq 1
\draw (1,-4) node[tre] (2) {}; \etq 2
\draw (3,-4) node[tre] (3) {}; \etq 3
\draw (4,-4) node[tre] (4) {}; \etq 4
\draw [->](r)--(a);
\draw [->](r)--(c);
\draw [->](a)--(1);
\draw [->](a)--(A);
\draw [->](a)--(B);
\draw [->](b)--(A);
\draw [->](b)--(B);
\draw [->](c)--(b);
\draw [->](c)--(4);
\draw [->](A)--(2);
\draw [->](B)--(3);
\end{tikzpicture}
\end{center}
\caption{\label{fig:contr6} Two non-isomorphic reduced phylogenetic networks at $\mu$-distance 0.}
\end{figure}

\begin{remark}
\label{rem1}
It is important to point out that  the $\mu$-distance does not satisfy the separation axiom on the class of reduced phylogenetic networks: for instance, the reduced networks $N_9$ and $N_{10}$ in \cite[Fig.~11]{cardona.ea:07a}, which we recall in Fig.~\ref{fig:contr6}, have the same $\mu$-representations, but they are not isomorphic. Therefore, Nakhleh's $m$ metric has a stronger separating power than the $\mu$-distance, in the sense that it satisfies the separation axiom in every class where $d_\mu$ satisfies it, and in at least one class where $d_\mu$ does not satisfy it.
\end{remark}

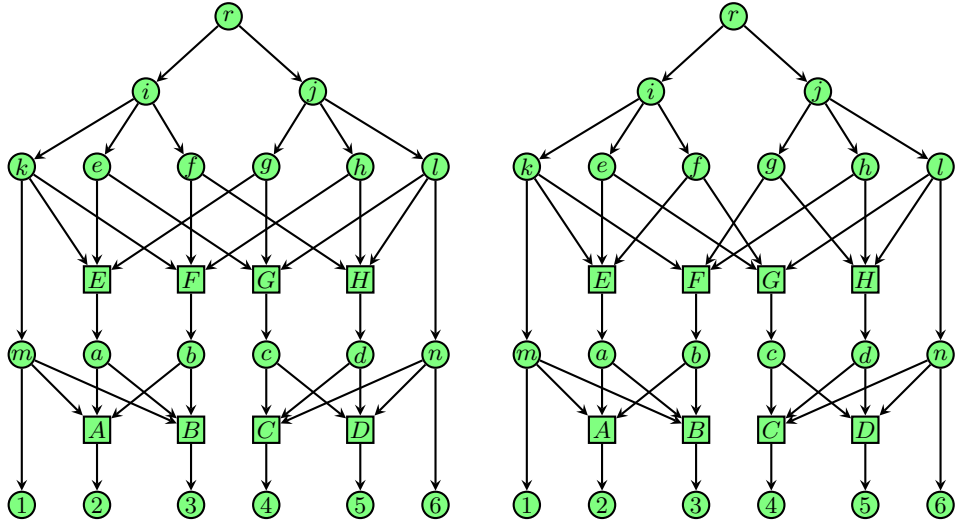
\begin{figure}[htb]
\begin{center}
\begin{tikzpicture}[scale=1,thick,>=stealth]
\draw(0,0) node[tre] (2) {}; \etq 2
\draw(1.25,0) node[tre] (3) {}; \etq 3
\draw(2.25,0) node[tre] (4) {}; \etq 4
\draw(3.5,0) node[tre] (5) {}; \etq 5
\draw(0,1) node[hyb] (A) {}; \etq A
\draw(1.25,1) node[hyb] (B) {}; \etq B
\draw(2.25,1) node[hyb] (C) {}; \etq C
\draw(3.5,1) node[hyb] (D) {}; \etq D
\draw(0,2) node[tre] (a) {}; \etq a
\draw(1.25,2) node[tre] (b) {}; \etq b
\draw(2.25,2) node[tre] (c) {}; \etq c
\draw(3.5,2) node[tre] (d) {}; \etq d
\draw(0,3) node[hyb] (E) {}; \etq E
\draw(1.25,3) node[hyb] (F) {}; \etq F
\draw(2.25,3) node[hyb] (G) {}; \etq G
\draw(3.5,3) node[hyb] (H) {}; \etq H
\draw(0,4.5) node[tre] (e) {}; \etq e
\draw(1.25,4.5) node[tre] (f) {}; \etq f
\draw(2.25,4.5) node[tre] (g) {}; \etq g
\draw(3.5,4.5) node[tre] (h) {}; \etq h
\draw(0.65,5.5) node[tre] (i) {}; \etq i
\draw(2.87,5.5) node[tre] (j) {}; \etq j
\draw(-1,4.5) node[tre] (k) {}; \etq k
\draw(4.5,4.5) node[tre] (l) {}; \etq l
\draw(-1,2) node[tre] (m) {}; \etq m
\draw(4.5,2) node[tre] (n) {}; \etq n
\draw(-1,0) node[tre] (1) {}; \etq 1
\draw(4.5,0) node[tre] (6) {}; \etq 6
\draw(1.75,6.5) node[tre] (r) {}; \etq r
 \draw[->](i)--(k);
 \draw[->](j)--(l);
    \draw[->](k)--(E);
    \draw[->](k)--(F);
    \draw[->](k)--(m);
    \draw[->](l)--(G);
    \draw[->](l)--(H);
    \draw[->](l)--(n);
    \draw[->](m)--(A);
    \draw[->](m)--(B);
  \draw[->](m)--(1);
    \draw[->](n)--(C);
    \draw[->](n)--(D);
    \draw[->](n)--(6);
 \draw[->](r)--(i);
 \draw[->](r)--(j);
 \draw[->](i)--(e);
    \draw[->](i)--(f);
    \draw[->](j)--(g);
    \draw[->](j)--(h);
    \draw[->](e)--(E);
    \draw[->](f)--(F);
    \draw[->](g)--(G);
    \draw[->](h)--(H);
    \draw[->](e)--(G);
    \draw[->](f)--(H);
    \draw[->](g)--(E);
    \draw[->](h)--(F);
    \draw[->](E)--(a);
    \draw[->](F)--(b);
    \draw[->](G)--(c);
    \draw[->](H)--(d);
    \draw[->](a)--(A);
    \draw[->](a)--(B);
    \draw[->](b)--(A);
    \draw[->](b)--(B);
    \draw[->](c)--(C);
    \draw[->](c)--(D);
    \draw[->](d)--(C);
    \draw[->](d)--(D);
    \draw[->](A)--(2);
    \draw[->](B)--(3);
    \draw[->](C)--(4);
    \draw[->](D)--(5);
  \end{tikzpicture}
  \qquad
    \begin{tikzpicture}[scale=1,thick,>=stealth]
\draw(0,0) node[tre] (2) {}; \etq 2
\draw(1.25,0) node[tre] (3) {}; \etq 3
\draw(2.25,0) node[tre] (4) {}; \etq 4
\draw(3.5,0) node[tre] (5) {}; \etq 5
\draw(0,1) node[hyb] (A) {}; \etq A
\draw(1.25,1) node[hyb] (B) {}; \etq B
\draw(2.25,1) node[hyb] (C) {}; \etq C
\draw(3.5,1) node[hyb] (D) {}; \etq D
\draw(0,2) node[tre] (a) {}; \etq a
\draw(1.25,2) node[tre] (b) {}; \etq b
\draw(2.25,2) node[tre] (c) {}; \etq c
\draw(3.5,2) node[tre] (d) {}; \etq d
\draw(0,3) node[hyb] (E) {}; \etq E
\draw(1.25,3) node[hyb] (F) {}; \etq F
\draw(2.25,3) node[hyb] (G) {}; \etq G
\draw(3.5,3) node[hyb] (H) {}; \etq H
\draw(0,4.5) node[tre] (e) {}; \etq e
\draw(1.25,4.5) node[tre] (f) {}; \etq f
\draw(2.25,4.5) node[tre] (g) {}; \etq g
\draw(3.5,4.5) node[tre] (h) {}; \etq h
\draw(0.65,5.5) node[tre] (i) {}; \etq i
\draw(2.87,5.5) node[tre] (j) {}; \etq j
\draw(-1,4.5) node[tre] (k) {}; \etq k
\draw(4.5,4.5) node[tre] (l) {}; \etq l
\draw(-1,2) node[tre] (m) {}; \etq m
\draw(4.5,2) node[tre] (n) {}; \etq n
\draw(-1,0) node[tre] (1) {}; \etq 1
\draw(4.5,0) node[tre] (6) {}; \etq 6
\draw(1.75,6.5) node[tre] (r) {}; \etq r
 \draw[->](i)--(k);
 \draw[->](j)--(l);
    \draw[->](k)--(E);
    \draw[->](k)--(F);
    \draw[->](k)--(m);
    \draw[->](l)--(G);
    \draw[->](l)--(H);
    \draw[->](l)--(n);
    \draw[->](m)--(A);
    \draw[->](m)--(B);
  \draw[->](m)--(1);
    \draw[->](n)--(C);
    \draw[->](n)--(D);
    \draw[->](n)--(6);
 \draw[->](r)--(i);
 \draw[->](r)--(j);
 \draw[->](i)--(e);
    \draw[->](i)--(f);
    \draw[->](j)--(g);
    \draw[->](j)--(h);
    \draw[->](e)--(E);
    \draw[->](f)--(E);
    \draw[->](g)--(F);
    \draw[->](h)--(F);
    \draw[->](e)--(G);
    \draw[->](f)--(G);
    \draw[->](g)--(H);
    \draw[->](h)--(H);
    \draw[->](E)--(a);
    \draw[->](F)--(b);
    \draw[->](G)--(c);
    \draw[->](H)--(d);
    \draw[->](a)--(A);
    \draw[->](a)--(B);
    \draw[->](b)--(A);
    \draw[->](b)--(B);
    \draw[->](c)--(C);
    \draw[->](c)--(D);
    \draw[->](d)--(C);
    \draw[->](d)--(D);
    \draw[->](A)--(2);
    \draw[->](B)--(3);
    \draw[->](C)--(4);
    \draw[->](D)--(5);
  \end{tikzpicture}
\end{center}
\caption{\label{fig:contr1} These time consistent tree-sibling phylogenetic networks have the same
nested labels representations, but they are not
isomorphic}
\end{figure}

\begin{remark}
\label{rem2}
It is false in general that if two arbitrary time consistent tree-sibling phylogenetic networks
$N_1$ and $N_2$ on the same set $S$ of taxa are such that $m(N_1,N_2)=0$, then $N_1\cong N_2$.
For instance, it is easy to check that the networks depicted in
Fig.~\ref{fig:contr1} have the same nested labels representations, but they are
not isomorphic. Thus, Nakhleh's dissimilarity measure is not a metric on the class of all 
time consistent tree-sibling phylogenetic networks.
\end{remark}

\subsection{$m$ singles out MUL-trees}

The comparison of MUL-trees generalizes  simultaneously the comparison of non-labeled rooted trees (understood as MUL-trees with all their leaves labeled with the same label)  and of rooted phylogenetic trees (MUL-trees where each leaf has one label, and different leaves have different labels).
Ganapathy \textsl{et al} have recently proposed in \cite{ganapathy.ea:TCBB06} two metrics for MUL-trees, an edition distance that generalizes the Robinson-Foulds distance for phylogenetic trees, and a metric based on the computation of the multi-labeled analogous of a Maximum Agreement Subtree.
n this subsection we show that the natural generalization  of Nakhleh's $m$ to MUL-trees is also a metric on the space of MUL-trees on a given set $S$ of taxa.

The definition of  the nested labeling  generalizes to MUL-trees in a natural way as follows.

\begin{definition}
Let $M=(V,E)$ be a MUL-tree on a set $S$ of labels. The \emph{nested labeling} $\ell(v)$ of the
nodes $v$ of $N$ is defined by induction on $h(v)$ as follows:

\begin{itemize}
\item If $h(v)=0$, that is, if $v$ is a leaf, and if the set of labels of $v$ is $S_v\subseteq S$,  then $\ell(v)=S_v$.

\item If $h(v)=m>0$ and if  the children of $v$ are  $v_1,\ldots,v_k$,
then $\ell(v)$ is the multiset of their nested labels:
$$
\ell(v)=\{\ell(v_1),\ldots,\ell(v_k)\}.
$$
\end{itemize}
The \emph{nested labels representation} of $M$ is the multiset $\Upsilon(M)$ of nested labels of its nodes (where each nested label appears with
multiplicity the number of nodes having it).
\end{definition}

With this definition of nested labels, Nakhleh's dissimilarity measure $m$ for MUL-trees is simply defined as in Proposition \ref{prop:m-nl}: half the cardinal of the symmetric difference of the nested labels representations.

Notice now that, in a MUL-tree, the nested label of a node yields, after replacing brackets by parentheses, the Newick string of the subtree rooted at that node. In particular, if two nodes in two MUL-trees have the same nested labels, then the subtrees rooted at them are isomorphic.
This remark lies at the basis of the following proof.

\begin{proposition}
Let $M_1$ and $M_2$ be two MUL-sets on the same set $S$ of labels. If $\Upsilon(M_1)=\Upsilon(M_2)$, then  $M_1\cong M_2$.\qed
\end{proposition}

\begin{proof}
Let $\Upsilon(M_1)$ and $\Upsilon(M_2)$ be the nested labels representations of $M_1$ and $M_2$.
 If $m(M_1,M_2)=0$, then $\Upsilon(M_1)=\Upsilon(M_2)$. Now, 
if $r_1$ and $r_2$ are the roots of $M_1$ and $M_2$, respectively, then 
$\ell(r_1)$ and $\ell(r_2)$ are the nested labels with highest level of nesting in $\Upsilon(M_1)$ and $\Upsilon(M_2)$, respectively,  and then, these multisets being equal, it must happen that 
$\ell(r_1)=\ell(r_2)$. But then the subtrees of $M_1$ and $M_2$ rooted at $r_1$ and $r_2$, that is, the MUL-trees $M_1$ and $M_2$ themselves, are isomorphic.
\qed
\end{proof}

In particular, 
 $m$ is a metric on the class of all MUL-trees labeled in a given set $S$.

\section{Conclusion}
In this paper we have complemented Luay Nakhleh's latest proposal of a
metric $m$ for phylogenetic networks by  showing that it separates
distinguishable networks,  and  that it satifies the separation axiom on the classes of tree-child and of quasi binary time consistent tree-sibling phylogenetic networks as well as of area cladograms. When $m$   is applied to phylogenetic trees, it
yields half the symmetric differences of the sets of (isomorphism
classes of) subtrees \cite{nakhleh:2007}, and it can be computed in time  polynomial in the size  of the networks.

Given a set $S$ of $n\geq 2$ labels, there exists no upper bound for the values of
 $m(N_{1},N_{2})$,
as there exist arbitrarily large phylogenetic networks with $n$ 
leaves and no internal node of any one of them equivalent to an 
internal node of the other one.



\end{document}